\documentclass[twocolumn,showpacs,preprintnumbers,amsmath,nofootinbib,aps,superscriptaddress]{revtex4}

\usepackage[dvips]{graphicx}
\usepackage{dcolumn}
\usepackage{bm,xcolor}

\newcommand*{\no}{\noindent}
\newcommand*{\bea}{\begin{eqnarray}}
\newcommand*{\eea}{\end{eqnarray}}
\newcommand*{\be}{\begin{equation}}
\newcommand*{\ee}{\end{equation}}

\graphicspath{{./}{publPlots/}}

\begin{document}


\title{The phase diagram of a gauge theory with fermionic baryons}

\author{Axel Maas}
\email{axelmaas@web.de}
\affiliation{Theoretical-Physical Institute, Friedrich-Schiller-University Jena, Max-Wien-Platz 1, D-07743 Jena, Germany}
\author{Lorenz von Smekal}
\email{lorenz.smekal@physik.tu-darmstadt.de}
\affiliation{Institut  f\"{u}r Kernphysik, Technische Universit\"{a}t
  Darmstadt, D-64289 Darmstadt, Germany} 
\author{Bj\"orn Wellegehausen}
\email{bjoern.wellegehausen@uni-jena.de}
\affiliation{Theoretical-Physical Institute, Friedrich-Schiller-University Jena, Max-Wien-Platz 1, D-07743 Jena, Germany}
\author{Andreas Wipf}
\email{wipf@tpi.uni-jena.de}
\affiliation{Theoretical-Physical Institute, Friedrich-Schiller-University Jena, Max-Wien-Platz 1, D-07743 Jena, Germany}

\date{\today}

\begin{abstract}

The fermion-sign problem at finite density is a persisting challenge
for Monte-Carlo simulations. Theories that do not
have a sign problem can provide valuable guidance and insight for
physically more relevant ones that do. Replacing the gauge group SU(3)
of QCD by the exceptional group G$_\text{2}$, for example, leads to
such a theory. It has mesons as well as bosonic and fermionic baryons, 
and shares many features with QCD. This makes the G$_\text{2}$ gauge theory 
ideally suited to study general properties of dense,
strongly-interacting matter, including baryonic and nuclear Fermi
pressure effects. Here we present the first-ever results from
lattice simulations of G$_\text{2}$ QCD with dynamical fermions,
providing a first explorative look at the phase diagram of this QCD-like theory at finite
temperature and baryon chemical potential. 

\end{abstract}

\pacs{11.30.Rd 12.38.Aw 12.38.Gc 12.38.Mh 21.65.Qr}

\maketitle

Finite fermion density continues to be a serious challenge for
Monte-Carlo simulations due to the fermion-sign problem
\cite{Gattringer:2010zz,deForcrand:2010ys}. The sign problem appears in many areas of
physics, but is of notorious importance to dense quark systems,
especially in nuclei, heavy-ion collisions, and compact stellar
objects. An alternative are models and continuum methods which do not have
this type of problem
\cite{Leupold:2011zz,Buballa:2003qv,Pawlowski:2010ht,Braun:2011pp,Maas:2011se}.
 However, these usually require approximations, and cross checks through lattice simulations remain desirable to improve
systematic reliability. 

To provide support from numerical simulations, two major
strategies have been followed. One is to replace the baryon chemical 
potential by some quantity more amenable to simulations, e.g.\ imaginary \cite{Bonati:2012pe,deForcrand:2010he,Cea:2012ev} or isospin
\cite{Kogut:2004zg,deForcrand:2007uz} chemical potential. The other is to replace the
theory with one accessible through numerical simulations at finite
density. However, such theories usually differ from the
original one in more or less important aspects.

One very well studied replacement of QCD for strongly interacting
matter at finite density is two-color QCD
\cite{Kogut:2000ek,Hands:2000ei,Hands:2006ve,Hands:2011ye,Strodthoff:2011tz}. In
this case, the baryons are bosons instead of fermions, however. This 
leads to profound differences, such as Bose-Einstein condensation of
a baryon superfluid with a BEC-BCS crossover at high densities instead
of the usual liquid-gas transition of nuclear matter. While two-color
QCD has many interesting aspects that deserve to be studied in their
own right, the quantum effects due to the fermionic nature of baryons
are expected to play a very significant role for nuclear matter and
especially in the physics of compact stellar objects
\cite{BraunMunzinger:2008tz}.      

Therefore, a more realistic replacement theory in this regard should
contain fermionic baryons. One possibility is the strong-coupling
limit \cite{deForcrand:2009dh}. In order to maintain the connection
with the continuum, however, we employ
here a different theory without sign problem for Monte-Carlo simulations. 
It is obtained by replacing the SU(3) gauge group of QCD with the
gauge group G$_\text{2}$ \cite{Holland:2003jy}. All color representations of
this theory are equivalent to real ones. As a consequence the Dirac
operator has an anti-unitary symplectic symmetry which for $N_f$
flavors leads to an extended
Pauli-G\"ursey SU(2$N_f)\times $Z$_2$ flavor symmetry
\cite{Kogut:2000ek,Holland:2003jy}, and is thus expected to have a non-anomalous
component even for a single flavor. In this paper we study the phase
diagram for a single Dirac flavor of Wilson fermions, corresponding to
a continuum SU(2)$\times$Z$_2$ extended flavor symmetry in the
chiral limit. Spontaneous or explicit breaking reduces this to 
U(1)$\times$Z$_2$ \cite{Maas:2012aa}. 
The unbroken U(1) relates to the baryon number to which the baryon
chemical potential is coupled. 
The anti-unitary symplectic symmetry of the Dirac
operator in a real color representation 
implies a two-fold degeneracy of its eigenvalues and hence 
that the fermion determinant remains positive at
finite baryon chemical potential even for a single flavor
\cite{Kogut:2000ek,Hands:2000ei}.  

The physical bound states of this theory, besides the
usual quark-antiquark and three-quark states, also contain hybrids of
one quark with three gluons, as well as diquarks and further bound
states with more than three quarks \cite{Holland:2003jy}. Thus, the
hadronic spectrum contains both fermionic as well as bosonic baryons 
and mesons. These bound states are created by very similar
interactions as in QCD, i.\ e.\ by a potential which rises linearly with
the separation of the quarks before string-breaking sets in
\cite{Holland:2003jy,Liptak:2008gx,Greensite:2006sm,Pepe:2006er,Wellegehausen:2010ai}. In 
fact, in a simple quark model picture with light quarks, one has a baryon-mass hierarchy
where for a sufficiently small quark mass nucleonic three-quark states
will be the second lightest baryons, above the light diquarks
as the would-be-Goldstone bosons of the extended flavor-symmetry
breaking. Hybrids, tetraquarks and other baryonic multi-quark states are expected to be much heavier.  

Another appealing aspect of the G$_\text{2}$ gauge group is that it
might be deformable to ordinary QCD by breaking  G$_\text{2}$ down
to SU(3) via a Higgs mechanism
\cite{Holland:2003jy,Wellegehausen:2011sc}, although this will likely
require several Higgs fields and various Yukawa couplings, CKM-type
explicit flavor violations, and further effects. The surplus bound states would then become 
heavy and disappear from the spectrum. The sign problem would
emerge, likewise. If it can be controlled
by the strength of the breaking, however, this could provide new
insights, a possibility that certainly
deserves further study in the future.   

For our simulations, which consumed roughly 2000 core-years, we employ an extension
of the available local HMC algorithm for scalars
\cite{Wellegehausen:2011sc}, generalized from QCD simulations, for
details see \cite{Wellegehausen:2011jz,Maas:2012aa}. The
introduction of temperature and chemical potential proceeds as in QCD. Our aim here is a first exploration of the phase diagram as a
proof of principle. One obstacle is the presence of a bulk transition
\cite{Pepe:2006er,Cossu:2007dk,Wellegehausen:2011sc}, which turns out to persist with
dynamical fermions  
\cite{Maas:2012aa,Wellegehausen:2011jz}. To avoid this we 
chose a lattice with at least $N_t=5$ time slices in finite
temperature simulations. Simulations of this theory incur considerable
 computational costs so that we restricted our lattices to $N_s=16$
 points in spatial directions. Altogether, we investigated three
 different sets of lattice parameters:

\begin{figure}
\includegraphics[width=0.5\linewidth]{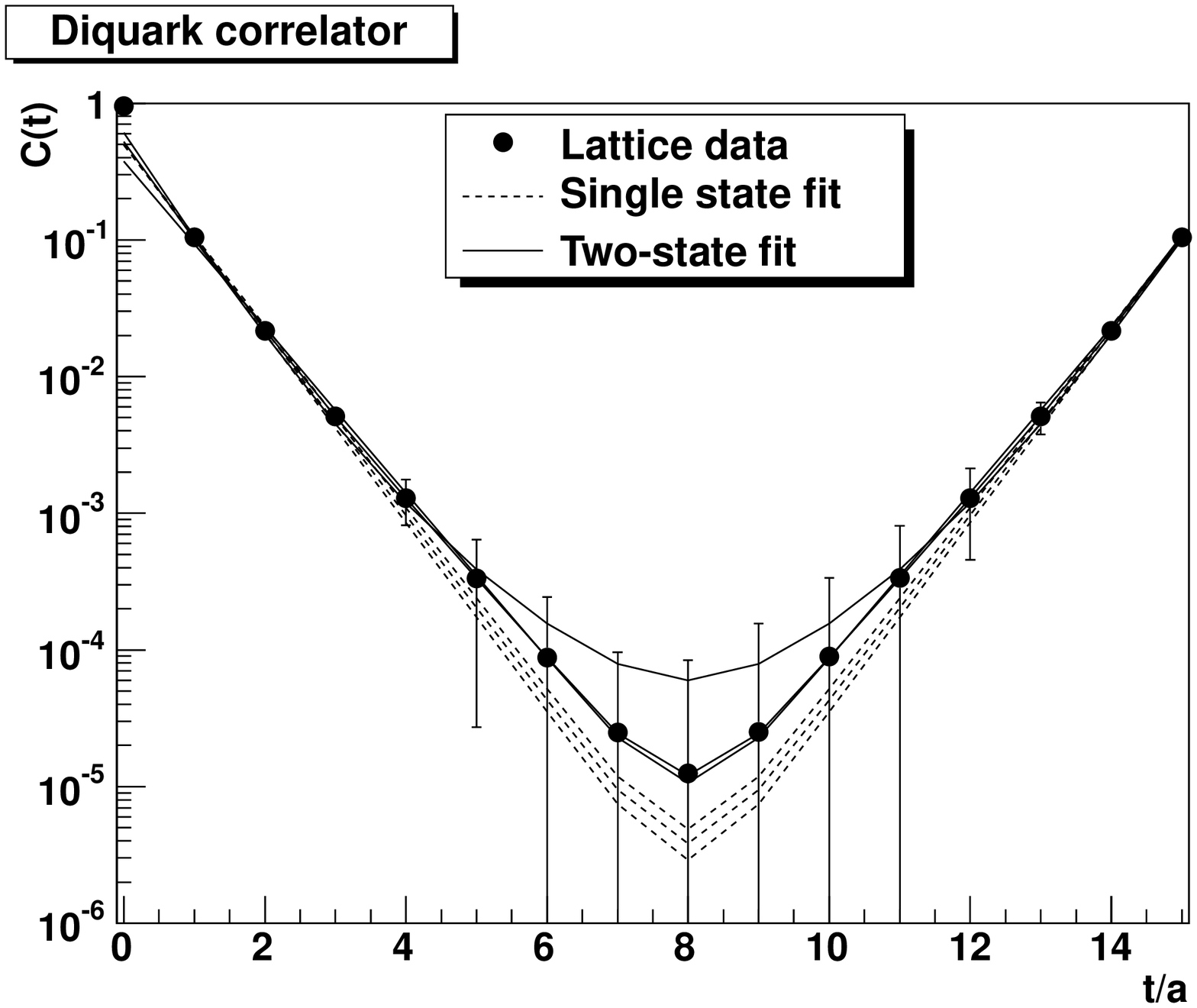}\includegraphics[width=0.5\linewidth]{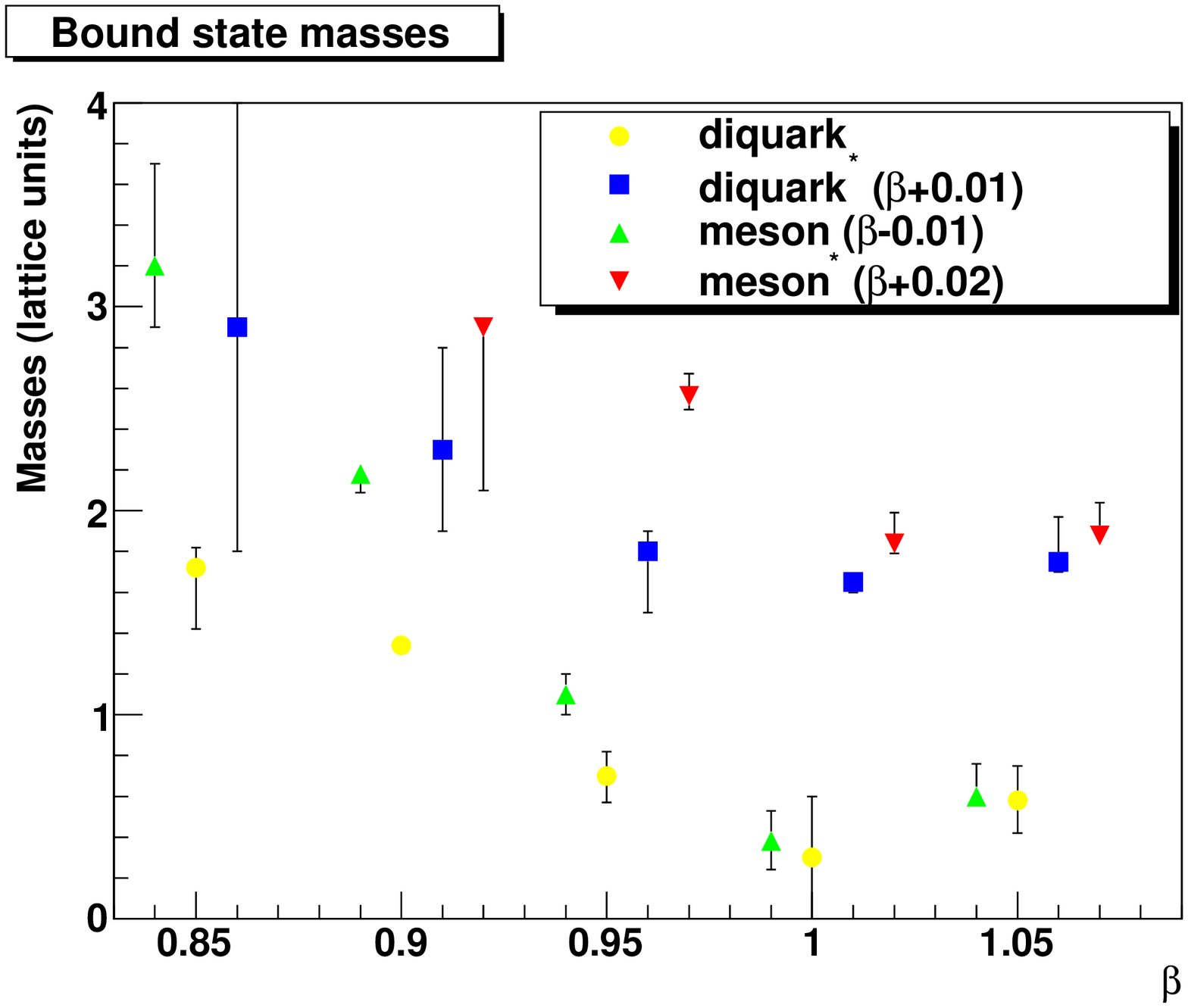}
\caption{The left panel shows an example for the mass
  correlator fits performed. Since in most cases no clear mass
  plateaus appear due to closeness of the excited states, conservative
  error estimates have been used, which are shown in the example as
  error bands. The right panel shows the masses obtained in this
  way. Note the shifted $\beta$ values to disentangle the plot. All
  results are for $\kappa=0.15625$. Only the range $\beta=0.9$ to
  $\beta=1.0$ has been used in the phase diagram calculations.}
\label{fig-mass} 

\vspace*{-.4cm}
\end{figure}

\no (a) At zero density we varied the lattice coupling $\beta$ between 0.9
and 1.0 in order to control temperature on our $5\times 16^3$ and $6\times 16^3$
lattices.
 
\no (b) At finite chemical potential $\mu$ we used $\beta=0.9$ on both,
$6\times 16^3$ and $8\times 16^3$ lattices, and $\beta = 1.0$ in
zero-temperature simulations on a $16^4$ lattice. The hopping
parameter in the fermion determinant was fixed at $\kappa=0.15625$ in
all these thermodynamic simulations. 

\no (c) For comparison we also varied 
$\kappa$ from $0.15385$ to $0.15625$ on the symmetric $16^4$
lattice with $\mu =0$ and $\beta $ in $0.85$ to $1.1$.
\cite{Maas:2012aa}.  

More results from smaller lattices to assess systematic effects will be reported elsewhere.

We measured three observables to study the phase diagram. One
is the Polyakov loop. Unlike QCD, due to the trivial center,
it is not an order parameter of the quenched theory
\cite{Holland:2003jy}, but it nevertheless reflects the corresponding
first-order phase transition very well
\cite{Pepe:2006er,Cossu:2007dk}. In fact, it remains so small in the
low-temperature phase that it is only possible to
determine upper bounds. We found this to be true also with dynamical
quarks. The second observable is the chiral condensate. The quenched
 G$_\text{2}$ theory has only one first-order transition at finite
 temperature which manifests itself also in the chiral condensate
\cite{Danzer:2008bk}, a feature that it shares with QCD and
two-color QCD. This is in contrast to QCD with adjoint quarks, where there is no sign problem either \cite{Kogut:2000ek},
but where separate chiral and deconfinement transitions occur
at largely different temperatures
\cite{Karsch:1998qj,Engels:2005te,Bilgici:2009jy}. 
We normalize the chiral condensate to its ($\beta$-dependent) vacuum
value, to avoid explicit renormalization. The
third observable is the baryon number density, the derivative of the
partition function with respect to the chemical potential. 
At large chemical potentials the density saturates to a
temperature-independent value. This is observed also in two-color QCD
\cite{Hands:2010vw}. 
To eliminate the scale, we therefore normalize the baryon density to its
saturation value.

Saturation on finite lattices can occur due to the Pauli principle.
In two-color QCD, this happens when all sites are occupied by bosonic
baryons such that the lattice has the maximal filling of $4N_f$ quarks per
site as allowed by the Pauli principle \cite{Hands:2006ve}. Here, we
also find the analogous saturation at a maximum of 14 quarks per
lattice site for the two spin states of a single quark flavor in the 7
dimenisonal fundamental color representation of the $G_2$ gauge
group. However, the fermion dynamics does not fully freeze out because
the plaquette average at saturation density is 0.5647(1), which is
smaller than the corresponding value of 0.5724(5) for the
pure gauge theory. 

\begin{figure}
\includegraphics[width=\linewidth]{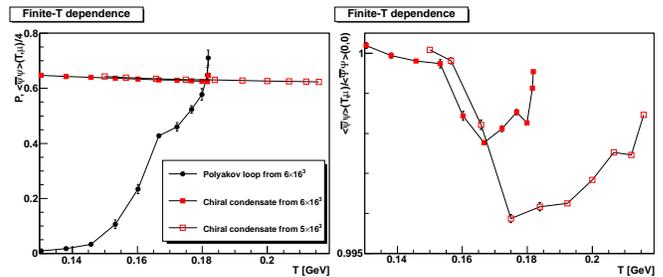}\\
\caption{The Polyakov loop and the unrenormalized chiral condensate at
  zero chemical potential (left panel) and the renormalized chiral
  condensate (right panel).}  
\label{fig-t}

\vspace*{-.4cm}
\end{figure}

\begin{figure*}
\includegraphics[width=0.85\linewidth]{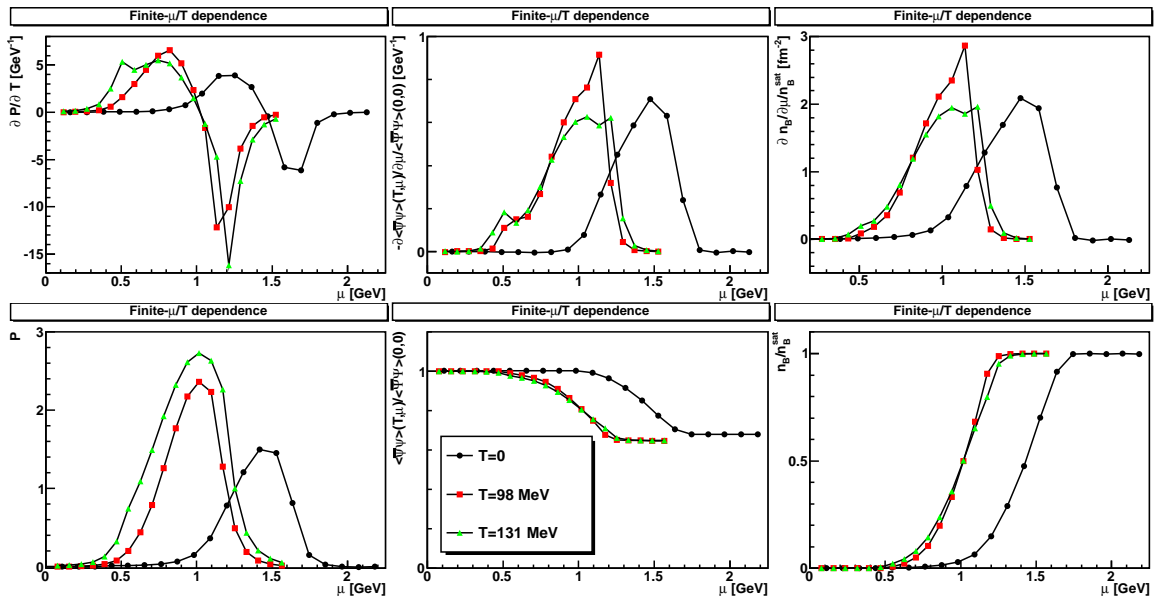}
\caption{Bottom row: raw Polyakov loop data (left),
  chiral condensate (middle), and normalized baryon density (right) at
  finite quark chemical potential $\mu $ and temperatures
  $T=0,\,98\,$MeV and $131\,$MeV. Top row: corresponding numerical derivatives.
}\label{fig-mu}

\vspace*{-.4cm}
\end{figure*}

In order to assess the phase diagram it is necessary to fix at least a
relative scale. Using bound state masses for this purpose is rather
challenging. We have determined the masses of the lowest
states in the scalar meson and diquark channels on the $16^4$ lattice,
see Figure \ref{fig-mass}. These include only the connected
contributions, however. The true scalar meson in our one-flavor case
contains disconnected contributions as well which will cause the
dominant splitting of its mass from that of the light diquark. 
The results show strong systematic effects \cite{Maas:2012aa}. 
Moreover, the diquarks are most sensitive to the quark mass 
due to their would-be-Goldstone nature \cite{vonSmekal:2012vx}. In
order to fix the scale from these masses would require a much more careful
and expensive mapping of the lines of constant physics in $\beta$ and
$\kappa$. Here we simply use the first excited state in the diquark
channel as the signal with the least sensitivity to the lattice parameters. 
It varies for
$\beta=0.9...1.0$ at $\kappa=0.15625$ from $2.3_{-4}^{+5}$ to
$1.65^{+1}_{-5}$ in lattice units. 
Requiring that the critical temperature for $\mu=0$ is at 160 MeV then 
leads to lattice spacings between
$0.25^{+5}_{-4}$ fm to $0.181^{+1}_{-20}$ fm in the same range for
$\beta $, with an acceptable systematic uncertainty for this first
exploratory investigation.   

The results at zero chemical potential are shown in Figure
\ref{fig-t}. The normalized chiral condensate only shows a rather weak
response to the transition observed in the Polyakov loop. In fact,
after a slight drop the chiral condensate starts rising again as 
temperature is further increased. That this behavior is likely to be a
lattice artifact is seen in particular by comparing the data from the $5 \times
16^3$ and $6 \times 16^3$ lattices as shown for the same range of 
couplings $\beta =0.9...1.0$ at $\kappa=0.15625$ in Fig.~\ref{fig-t}.

On one hand, this apparently unphysical behavior of the condensate
might be due to finite volume effects, because the three-dimensional volume
shrinks with increasing temperature when controlled by the lattice
coupling as it is done here. On the other hand, in the zero
temperature simulations we observe that the meson and diquark masses
at fixed $\kappa=0.15625$ start increasing with $\beta$ again just above
$\beta\approx 1$, c.f. Fig.~\ref{fig-mass}. This might indicate that we
encounter a transition into an unphysical phase such as the Aoki phase
\cite{Maas:2012aa}. Although we have not carefully mapped out the
lines of constant physics, for the highest temperatures in
Fig.~\ref{fig-t}, which correspond to $\beta = 1$, we might be getting
too close to or even into this unphysical phase. Both, finite-volume
effects and Aoki phase would lead to an unphysical
behavior as observed, a weakening of the chiral transition together with an
increase of the chiral condensate towards higher temperatures. 

While this implies that even larger and finer lattices will be
required to reliably quantify the systematic uncertainties in the high
temperature regime, we avoid these lattice artifacts in our finite
density simulations below as much as possible with 
controlling temperature by varying the number of time slices at
reasonably small $\beta $ values. From our present analysis we are
confident that the lattice parameters (b) used at finite chemical
potential, as described above, should be reasonably safe. A more
detailed quantitative analysis, in particular at temperatures near the
$\mu=0$ transition in Fig.~\ref{fig-t} and above, seems prohibitively
expensive at the moment.  

As a final comment on the $\mu = 0$ transition before we discuss our
finite density results we note that, strictly speaking,
there is no chiral symmetry in the usual sense in our one-flavor
theory. Rather, the condensate here measures the breaking of the
extended SU(2) Pauli-G\"ursey symmetry down to the usual baryon number
U(1). Thus, our intuition about the usual chiral restoration at the
deconfinement transition might not be a good guide in this case
either. It is quite possible, for example, that the 2-flavor theory 
behaves differently because it includes a standard chiral SU(2)$\times $SU(2)
component in its extended flavor SU(4) which is expected to be
broken at low temperatures and which might therefore show a stronger response to
the $\mu = 0$ transition in $G_2$ as well.   

\begin{figure*}
\includegraphics[width=0.85\linewidth]{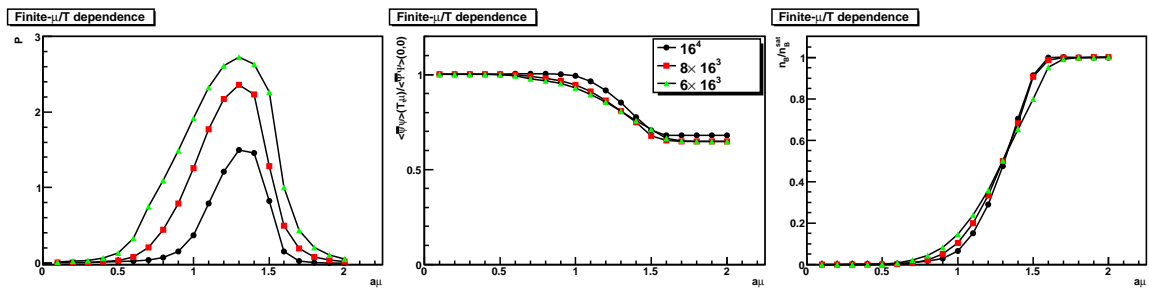}
\caption{The data from the bottom row of Figure
  \ref{fig-mu} in lattice units.}\label{fig-mu-lu}

\vspace*{-.4cm}
\end{figure*}

The behavior of Polyakov loop, chiral condensate and quark density at
finite chemical potential with the scale set as explained
above and in lattice units are shown in Figures \ref{fig-mu} and
\ref{fig-mu-lu}. At zero temperature the Polyakov loop peaks at around
1.5 GeV. Chiral condensate and baryon density both show a
rather broad transition near the maximum in the derivative of Polyakov
loop. At large chemical potential the density saturates, as 
discussed above. This effectively quenches the
theory which explains the decrease of the
Polyakov loop as in two-color QCD \cite{Hands:2006ve,Hands:2010vw}. It also
implies, however, that the regime beyond the peak in the Polyakov loop 
is affected by lattice artifacts thus limiting the chemical potential
to below 1.5 GeV at $T=0$ and to below about 1 GeV 
at the two finite temperatures considered here. 
This is emphasized by the result in lattice units where the saturation
is seen to be nearly independent of the lattice coupling.  

In the zero temperature results for the quark density one should see
the onset of Bose-Einstein condensation of diquarks when the quark
chemical potential reaches half the diquark mass. With a mass of the
lightest diquark around
$300-400$ MeV  here, this so-called silver-blaze point \cite{Cohen:2003kd} 
is thus expected around $\mu\approx 150-200$ MeV. On
smaller $16\times 8^3$ lattices, with larger masses and with
considarably better statistics, one indeed observes this onset of
diquark condensation in the baryon density which increases from zero
to a small finite plateau in a narrow region \cite{Maas:2012aa}. The
critical chemical potentials for this onset furthermore agree very
well with half the diquark masses as extracted from their 
correlators for a wide range of lattice parameters 
\cite{Maas:2012aa,Maas:2012ts}. It is a small effect, however. So
small that it cannot be seen on the scales of the Figures presented
here. A more detailed study with finer resolution on the present lattices 
requires much higher statistics. Having said that, the onset of the
large increase in density observed here, for $T=0$ at $\mu\approx 1 $ GeV, is
by the same argument unlikely to be due to these light pseudo-Goldstone
diquarks.  The first excited diquark states of
Fig.~\ref{fig-mass} should come in at $\mu $ around $400 -550$ MeV and 
thus still tend to be too small to explain this threshold.
The intriguing alternative would be that we do see the effect of
fermionic baryons here. If the quark-gluon hybrids and other
baryons are too heavy as expected, one would conclude that the density
in the region around the increase at  $\mu\approx 1 $ GeV is dominated
by three-quark states, the $G_2$ nucleons. We will find out with
further baryon spectroscopy in the future.

\begin{figure*}
\includegraphics[width=0.8\linewidth]{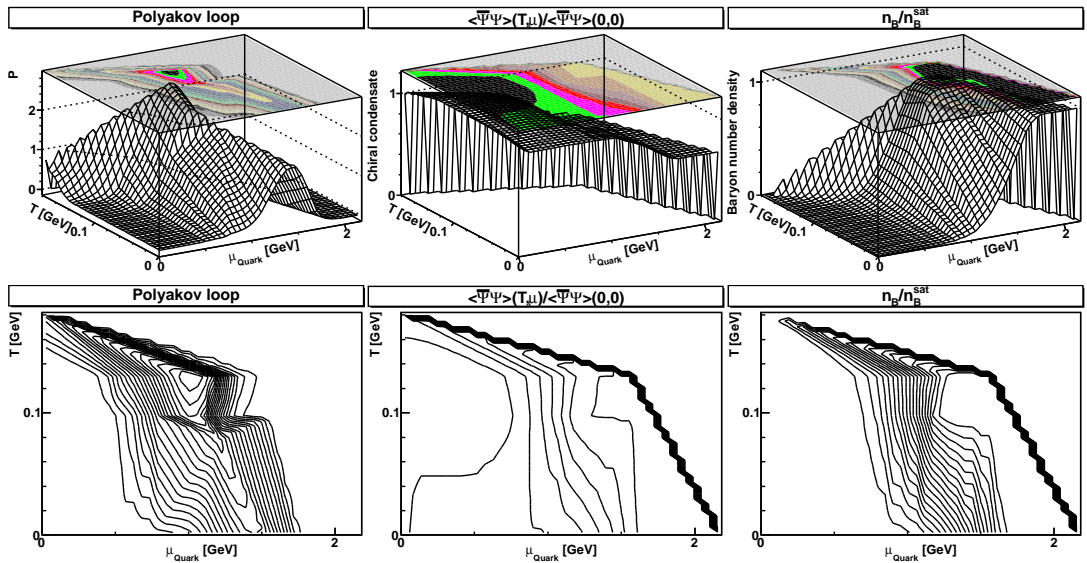}
\caption{\label{fig-pd} Interpolated phase diagram for Polyakov loop
  (left), chiral condensate (middle), and baryon density (right).}

\vspace*{-.4cm}
\end{figure*}

Finally, when both  temperature and chemical potential 
are non-zero, a gradual shift towards smaller chemical potentials along with a
broadening of the transition is observed. 
The interpolated phase diagram in Figure \ref{fig-pd} exhibits a
rather polygonal shape but is potentially affected by systematic
effects in the scale setting. Results on smaller lattices show the
same general features albeit with larger systematic errors
\cite{Maas:2012aa}.

In total, the phase diagram of $G_2$ QCD shows a number of
interesting features which deserve further systematic investigation to
see whether they are genuinely physical, or whether some of them are
lattice artifacts. Given sufficient computational resources its complete
phase diagram can be mapped with any desired precision. In particular,
the question whether one can have a chiral first-order transition
rather than a smooth crossover at finite density is very significant
as it would imply the presence of a critical endpoint in a theory with
properties so close to the ones of QCD, where this is of prime
importance for future heavy-ion experiments.  
 
To summarize, we have determined for the first time, but still at an
exploratory level, the phase diagram of G$_\text{2}$ QCD, a
non-Abelian gauge theory with fermionic baryons. This opens new avenues
for high density studies of the strong force from first-principles
calculations. In particular, it will be possible to assess the
role of the Fermi statistics at finite baryon density which is of 
great importance, e.g. for compact stellar objects. One might ask at 
which relative densities quark or hadron equations of state are more
favorable, and thus whether only neutron stars or also
quark stars can exist. Given the inflow of new astrophysical
observational data on compact stellar objects, including major new
discoveries, any insight on a fundamental level would be timely and
important. Of course, G$_\text{2}$ QCD is not QCD, and
such insight can only be qualitative. 

It also has to be assessed carefully  to what extent the particle
spectra of the two differ and what effects this has. However, these
ab-initio results provide a novel approach to test investigations of  
the true QCD phase diagram by effective models and functional
methods. If G$_\text{2}$ QCD is continuously deformable to QCD
\cite{Holland:2003jy,Wellegehausen:2011sc,Maas:2012aa}, 
this may proceed along the lines developed for Yang-Mills theory
\cite{Pawlowski:2010ht,Maas:2011se,Braun:2011pp}: Systematically check
approximations and model assumptions in functional methods along
this deformation from lattice calculations as long as possible, so that the
latter can then be used reliably in the final step towards 
QCD when the sign problem strikes back.

\medskip

\no{\bf Acknowledgments}

We are grateful to Simon Hands and Uwe-Jens Wiese for helpful
discussions and comments. B.W.\ was supported by the DFG graduate
school 1523-1,  
A.M.\ under DFG grant number MA 3935/5-1, A.W.\ under 
DFG grant number Wi 777/11, and L.v.S.\ by the Helmholtz
International Center for FAIR within the LOEWE program of the State of
Hesse, the Helmholtz Association Grant VH-NG-332, and the European
Commission, FP7-PEOPLE-2009-RG No.\ 249203. 
Simulations were performed on the LOEWE-CSC at the University
of Frankfurt and on the HPC cluster at the University of Jena.



\end{document}